\documentclass{article}

\usepackage[english]{babel}
\usepackage[letterpaper,top=2cm,bottom=2cm,left=3cm,right=3cm,marginparwidth=1.75cm]{geometry}
\usepackage{mathptmx}
\usepackage{sectsty}
\usepackage{amsmath}
\usepackage{bm}
\usepackage{graphicx} 
\usepackage{hyperref}

\usepackage{subcaption}

\usepackage[format=plain,justification=justified,singlelinecheck=false,font={stretch=1.125,small,sf},labelfont=bf,labelsep=space]{caption}

\usepackage[T1]{fontenc}
\usepackage[version=4]{mhchem}

\usepackage{newtxmath}
\usepackage{siunitx}
\usepackage{pdfpages}

\title{
	Molecular Nitrogen Formation in Nitrogen-Implanted (100) \ce{\beta-Ga2O3} Revealed by Temperature-Dependent N~$K$-edge XANES
}

\author{
    I.N.~Demchenko\textit{$^{a,b}$}$^{\ast}$, 
    Y.~Syryanyy\textit{$^{c}$}, 
    A.~Shokri\textit{$^{b}$}, 
    Y.~Melikhov\textit{$^{d}$}, 
    M.~Chernyshova\textit{$^{e}$}, \\
    M.~Turek\textit{$^{f}$}, 
    A.~Dro\'{z}dziel\textit{$^{f}$}, 
    F.~Munnik\textit{$^{g}$}, 
    R.~Jakie\l{}a\textit{$^{h}$}, 
    R.~Minikayev\textit{$^{h}$}, \\
    J.Z.~Domagala\textit{$^{h}$}, 
    A.~Derkachova\textit{$^{h}$}, 
    M.~Zaj\k{a}c\textit{$^{i}$}, 
    J.~Krajczewski\textit{$^{j}$}, \\
    E.~Grzanka\textit{$^{k}$}, 
    Z.~ Galazka\textit{$^{l}$} 
}

\begin{document}

\renewcommand{\thefootnote}{\fnsymbol{footnote}}
\renewcommand\footnoterule{\vspace*{1pt}
 \hrule width 3.5in height 0.4pt \vspace*{5pt}} 
\newcommand{\bee}{\begin{eqnarray}}
\newcommand{\eee}{\end{eqnarray}}
\newcommand{\ba}{\begin{array}}
\newcommand{\ea}{\end{array}}
\newcommand{\bt}{\begin{tabular}}
\newcommand{\et}{\end{tabular}}

\sisetup{
	range-phrase = --,
	range-units  = single
}

\maketitle

\begin{abstract}
The realization of $p$-type doping in wide-band-gap oxide semiconductors
remains a major challenge, particularly in
\ce{\beta-Ga2O3} where nitrogen has long been
considered a potential acceptor dopant but has consistently failed to
produce hole conductivity. Here we investigate the microscopic
configuration of implanted nitrogen in
(100)~\ce{\beta-Ga2O3} using
temperature-dependent N~$K$-edge x-ray absorption spectroscopy. The
spectra reveal a pronounced $\pi^*$ resonance characteristic of molecular
nitrogen, which becomes increasingly dominant upon thermal annealing.
First-principles calculations and multiple-scattering simulations reveal
a pronounced tendency for nitrogen atoms to form \ce{N-N} bonded
configurations in the \ce{Ga2O3} matrix,
particularly in defect-rich environments created by ion implantation,
reproducing the characteristic spectral features observed in the N
$K$-edge XANES spectra. Structural analysis further indicates that
implantation induces a defect-rich near-surface layer with local
\ce{\beta}→\ce{\gamma}-like structural motifs, highlighting the strongly nonequilibrium
structural environment in which nitrogen incorporation occurs. Reported
results show that implanted nitrogen preferentially forms molecular
\ce{N2}-like configurations rather than substitutional acceptors. Our results
provide a microscopic explanation for the long-standing failure of
nitrogen acceptor doping in \ce{\beta-Ga2O3} and
reveal dopant molecularization as a previously overlooked pathway for
impurity incorporation under strongly nonequilibrium implantation
conditions.
\end{abstract}

\footnotetext{\textit{$^{a}$}~The Centre for Advanced Materials and Technologies,
CEZAMAT at the Warsaw University of Technology, 19 Poleczki St, Warsaw 02-822, Poland}

\footnotetext{\textit{$^{b}$}~Institute of Plasma Physics and Laser Microfusion,
ul.~Hery 23, 01-497 Warsaw, Poland}

\footnotetext{\textit{$^{c}$}~Institute of Microelectronics and Optoelectronics,
Warsaw University of Technology, ul.~Koszykowa 75, 00-662 Warsaw, Poland}

\footnotetext{\textit{$^{d}$}~Institute of Fundamental Technological Research
Polish Academy of Sciences, ul.~Pawinskiego 5b, 02-106 Warsaw, Poland}

\footnotetext{\textit{$^{e}$}~National Center for Nuclear Research, Andrzeja
Sołtana 7, 05-400 Otwock, Poland}

\footnotetext{\textit{$^{f}$}~Institute of Physics, Maria Curie-Sklodowska
University, pl. M.~Curie-Skłodowskiej 1, 20-031 Lublin, Poland}

\footnotetext{\textit{$^{g}$}~Helmholtz-Zentrum Dresden-Rossendorf, Bautzner
Landstraße 400, 01328~Dresden, Germany}

\footnotetext{\textit{$^{h}$}~Institute of Physics PAS, al. Lotników 32/46, 02-668
Warsaw, Poland}

\footnotetext{\textit{$^{i}$}~National Synchrotron Radiation Centre SOLARIS,
Jagiellonian University, Czerwone Maki 98, 30-392 Kraków, Poland}

\footnotetext{\textit{$^{j}$}~Faculty of Chemistry, University of Warsaw,
Pasteura 1, 02-093 Warsaw, Poland}

\footnotetext{\textit{$^{k}$}~Institute of High Pressure Physics Polish Academy
of Sciences, ul. Sokolowska 29/37, 01-142 Warsaw, Poland}

\footnotetext{\textit{$^{l}$}~Leibniz-Institut für Kristallzüchtung,
Max-Born-Straße 2 12489 Berlin, Germany}

\footnotetext{$^{\ast}$~corresponding author: I.N.~Demchenko E-mail: iraida.demchenko@pw.edu.pl, iraida.demchenko@ifpilm.pl}

\vspace{0.5cm}

\ce{\beta-Ga2O3} has emerged as a prototypical
ultra-wide band gap (WBG) semiconductor for next-generation high-power
electronics owing to its large breakdown field, availability of bulk
substrates, and compatibility with scalable growth technologies
\cite{1,2,3,4}. Despite rapid advances in device architectures, a fundamental
materials limitation persists: the absence of reliable and stable $p$-type
conductivity \cite{5,6}. The inability to realize effective acceptor
doping in \ce{\beta-Ga2O3} constrains the
development of bipolar and complementary device concepts and remains one
of the central unresolved problems in this material system. A key
unresolved question is whether dopant atoms introduced under strongly
nonequilibrium conditions remain isolated within the lattice or instead
undergo spontaneous pairing or molecularization, thereby suppressing
their intended electronic activity.

Nitrogen in \ce{\beta-Ga2O3} provides a
particularly intriguing case. Despite being a long-considered acceptor
candidate due to its chemical similarity to oxygen and its role in other
oxide and nitride semiconductors, it has repeatedly failed to produce
hole conductivity. First-principles calculations predict that
substitutional nitrogen on oxygen sites (\ce{N_{O}}) forms deep acceptor
states \cite{7}, while various nitrogen-related complexes may act as
compensating centers or become energetically competitive depending on
Fermi-level position and growth conditions \cite{1,6}. In particular,
theoretical studies indicate that nitrogen can exhibit amphoteric
behavior and form multiple configurations, including complexes with
oxygen vacancies, which may substantially modify its electronic activity
\cite{2,6,7,8chapter5,8chapter6}.

Experimentally, nitrogen incorporation, whether during growth, plasma
exposure, or ion implantation, has consistently resulted in
semi-insulating behavior and deep defect levels rather than shallow
acceptor activation \cite{3,9,10}. Electrical measurements reveal strong
compensation and trap formation, yet the microscopic origin of this
self-compensation remains unresolved. Direct experimental determination
of the local bonding configuration of nitrogen in
\ce{\beta-Ga2O3}, therefore, remains scarce and
largely indirect.

Evidence from other wide-band-gap oxides suggests that the chemical
state of nitrogen may deviate significantly from the simple
substitutional acceptor picture. In ZnO, N~$K$-edge XANES studies have
demonstrated that nitrogen frequently forms molecular
\ce{N2}-like configurations within the lattice rather than
isolated substitutional \ce{N_{O}} centers, even when nominally
introduced as an acceptor dopant \cite{11,12}. 
The presence of characteristic $\pi^{*}$ resonances associated with \ce{N-N} bonding provided direct spectroscopic evidence for such molecular configurations. 
Moreover, Bazioti \textit{et al.}~\cite{13} using electron energy loss spectroscopy showed that after annealing, zinc vacancy clusters (\ce{V_{Zn}}) filled with \ce{N2} are formed, interpreted as evidence that nitrogen does not stabilize in the substitutional \ce{N_{O}} configuration and thereby limits $p$-type doping. 
These findings highlight that identical nominal dopant concentrations
can correspond to fundamentally different local bonding environments
with drastically different electronic consequences.

For \ce{\beta-Ga2O3}, N~$K$-edge XANES
investigations similarly indicate the coexistence of distinct nitrogen
species depending on processing conditions. Experimental spectra have
revealed features consistent with both Ga--N bonded states and
molecular-like nitrogen configurations, with the relative contribution
evolving under thermal treatment and nitridation conditions \cite{14}.
These observations suggest that nitrogen incorporation does not
necessarily yield a unique structural motif and that metastable or
molecular forms may be stabilized under quasi-equilibrium or defect-rich
environments.

The atomic configuration of nitrogen depends critically on the
incorporation pathway. While growth and high-temperature nitridation may
approach near-equilibrium bonding environments, ion implantation
represents a strongly non-equilibrium process, generating a dense
population of vacancies, interstitials and defect complexes. This
defect-rich environment fundamentally reshapes the thermodynamic
landscape, potentially stabilizing nitrogen configurations distinct from
isolated substitutional acceptors, including vacancy-assisted complexes
or \ce{N2}-like species \cite{2}. Consequently, the
microscopic mechanism governing nitrogen self-compensation under
implantation conditions remains experimentally unresolved. Specifically,
it remains unclear whether implanted nitrogen predominantly occupies
substitutional oxygen sites, forms vacancy-associated complexes, or
evolves toward molecular \ce{N2}-like configurations during post-implantation
thermal processing. Here we directly probe the local configuration of
implanted nitrogen using temperature-dependent N~$K$-edge XANES, aiming to
resolve the microscopic bonding environment responsible for
nitrogen-induced compensation in
\ce{\beta-Ga2O3}.

Crystal samples were prepared from a 2-inch diameter bulk \ce{\beta-Ga2O3} single crystal grown along the {[}010{]} crystallographic direction using the Czochralski method at the
Leibniz-Institut für Kristallzüchtung (Berlin, Germany). 
The crystal was grown from an Ir crucible with an oxygen concentration of \qty{8}{\percent} by volume in the growth atmosphere with no intentional dopants (for further details, see Refs.~\cite{15,16}).
The samples of size \qtyproduct{5 x 5 x 0.5}{\mm} were (100)-oriented and a double-sided chemical-mechanical polishing was performed.
The samples were semiconducting with the free electron concentration and electron mobility (from Hall effect measurements) of \qty{3.4e17}{\cm^{-3}} and \qty{118}{\cm^{2}.V^{-1}.s^{-1}}, respectively. 
Monocrystalline \ce{\beta-Ga2O3} samples were implanted with N ions of energy \qty{175}{\kilo\eV} at room temperature (RT). 
Base pressure in the irradiation chamber was about \qty{1e-6}{\milli\bar}. 
Irradiation currents' fluxes were of the order \qty{0.25}{\micro A/cm^{2}} and fluences were set to \qty{1e15}{}, \qty{5e15}{}, and \qty{1e16}{ions/cm^{2}}. 
A custom-made arc discharge ion source with an internal evaporator \cite{17,18} was used.
The implantation angle was set at approximately \ang{7} from the surface normal to avoid channeling.

X-ray absorption experiments were performed at the Solaris synchrotron on the PIRX beamline. 
The XANES spectra were obtained by recording the
total fluorescence yield (FLUO) signal from the samples while scanning
the photon energy across the N~$K$-edge region. The measurements were
carried out with the polarization vector of the synchrotron radiation
oriented close to the ``magic'' angle, i.e., the angle at which the
absorption cross section becomes independent of orbital orientation.
After normalization to the photon flux, the recorded N~$K$-edge XANES
spectra were subjected to subtraction of a linear background fitted to
the flat pre-edge region. For quantitative comparison, the spectra were
subsequently normalized to the atomic limit in the region approximately
\qty{30}{\eV} above the absorption edge, where angular dependence is negligible.
Note all XANES measurements were performed at RT on stepwise annealed
samples in the temperature range of \qtyrange[]{320}{1100}{\celsius}.

First-principles DFT calculations were performed using VASP (v. 6.3.2) that employs the projector-augmented wave method \cite{19,20,21,22}. 
The Heyd-Scuseria-Ernzerhof (HSE06) hybrid functional \cite{23} with the fraction of the Fock exchange $\alpha = 0.35$, and a screening parameter $\mu = \qty{0.20}{\angstrom^{-1}}$ was used. 
The pristine \ce{\beta-Ga2O3} has a monoclinic crystal structure with space group C2/m, with the lattice constants $a = \qty{12.23}{\angstrom}$, $b = \qty{3.04}{\angstrom}$, and $c = \qty{5.80}{\angstrom}$ \cite{7,24,25}. 
Defect formation energies and thermodynamic charge transition levels were calculated in a relaxed 160-atom \qtyproduct{1 x 4 x 2}{} supercell using standard formalism \cite{7,26,27}. 
Further simulation details are provided in \textbf{Supplemental Material}. 

The N~$K$-edge XANES spectra were calculated from the relaxed atomic structures using the FDMNES code \cite{28}. 
We primarily employed the computationally efficient multiple-scattering muffin-tin approximation
(MTA), which proved sufficient for atoms occupying regular lattice sites, including the Ga~$K$-edge and O~$K$-edge of \ce{Ga2O3} and the N~$K$-edge of \ce{Ga2O3}:N. 
Full-potential finite-difference (FDM) calculations confirmed that MTA yields very similar spectra that accurately reproduce bulk undisturbed \ce{Ga2O3} experimental data (not shown).

\begin{figure}[!h]
    \centering
    \includegraphics[width=1.0\textwidth]{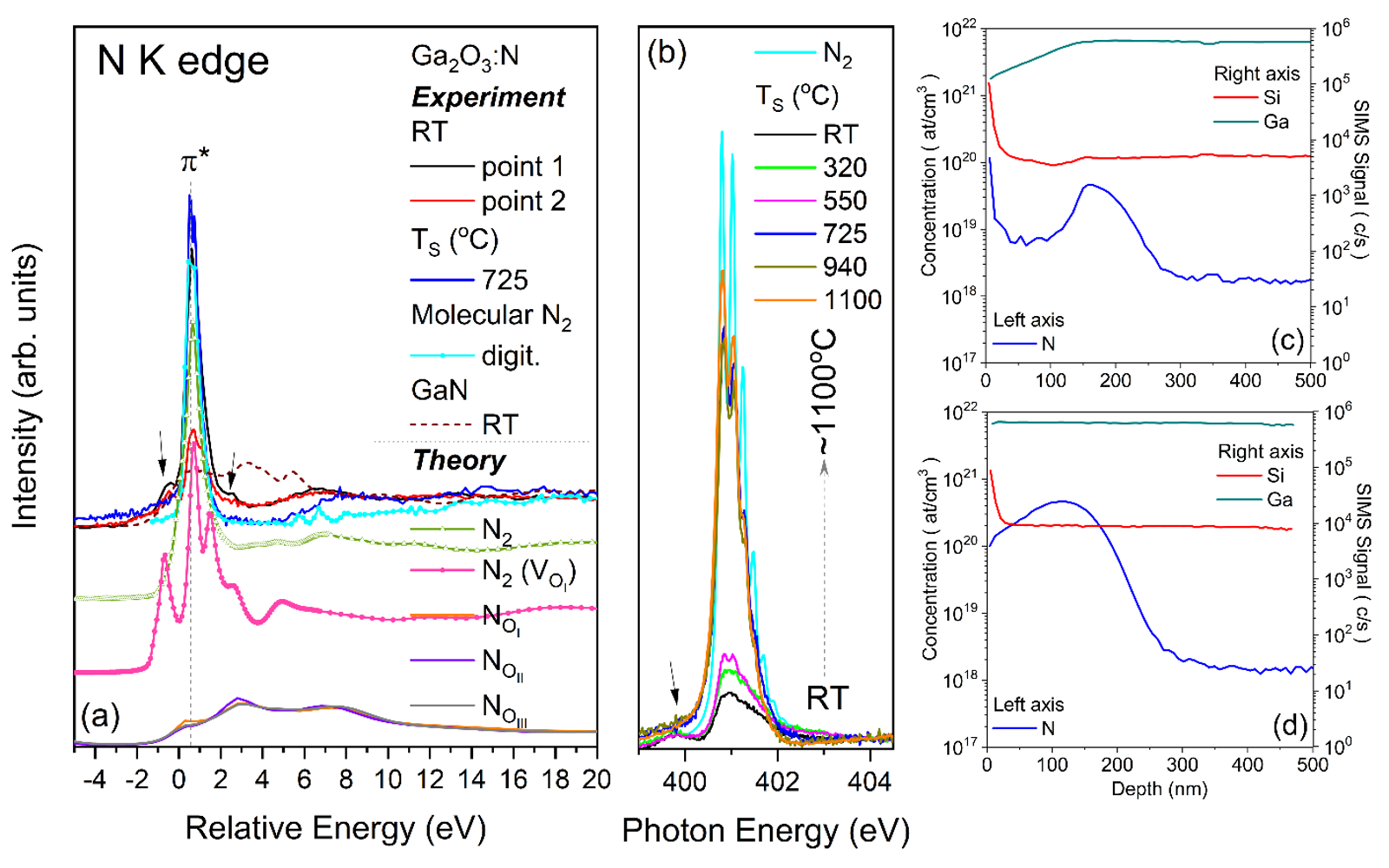}
    \caption{
    (a) Normalized N~$K$-edge XANES spectra of \ce{\beta-Ga2O3}:N implanted at room temperature and selected annealing at \qty{725}{\celsius} measured using FLUO detection mode. 
    The black/red/blue/cyan solid lines combined at the top represent the experimental spectra. 
    Spectrum corresponding to molecular nitrogen was digitized from \cite{34}. 
    Theoretical spectra calculated using FDMNES software were spread for clarity. 
    Models correspond to: three different substitutional \ce{N_{O}} (combined in down position), molecular nitrogen formed from two interstitial N (\ce{i5-i9}), and a similar molecular model in \ce{V_{O}}. 
    (b) High-resolution XANES spectra for implanted specimens in the whole range of annealing temperatures compared to the molecular nitrogen spectrum. 
    SIMS depth profiles of the investigated samples (c) implanted at RT and (d) subsequently annealed at \qty{1100}{\celsius}.     
    The nitrogen distribution maximum in the sample implanted at RT is in good agreement with SRIM calculations shown in Figure~\textbf{S3(b)}.
    }
    \label{fig:1}
\end{figure}

Figure~\ref{fig:1}~(a) presents the central experimental observation. 
All N-implanted \ce{\beta-Ga2O3} samples exhibit a pronounced near-edge resonance at the N~$K$-edge with a line shape dominated by a strong $\pi^{*}$ feature characteristic of \ce{N-N} bonding. 
The variation in $\pi^{*}$ resonance intensity between spectra acquired at
different surface locations (black vs. red curve) indicates pronounced
spatial inhomogeneity in the distribution of molecular nitrogen within
the implanted layer. 
The spectra are also clearly distinct from crystalline \ce{GaN} (brown dashed line), demonstrating that the implanted nitrogen does not predominantly form an extended \ce{Ga-N} network.

With increasing annealing temperature, the $\pi^{*}$ resonance systematically sharpens and gains spectral weight, while the overall near-edge profile evolves toward a more symmetric, molecular-like response (Figure~\ref{fig:1}(b)). 
This behavior indicates a thermally driven redistribution of nitrogen toward \ce{N2}-like configurations rather than activation into a substitutional \ce{N_{O}}-type state.

An additional spectral signature supporting this interpretation is the dip visible around \qty{\approx 4}{\eV} above the edge. 
This dependence is absent in the RT spectrum but develops after annealing at \qty{725}{\celsius} (blue line). 
At this temperature, the experimental spectrum closely resembles the reference spectrum of molecular \ce{N2} (cyan line). 
The simultaneous presence of the sharp $\pi^{*}$ resonance and the characteristic post-edge dip therefore provides strong spectroscopic evidence for the formation of molecular nitrogen species within the implanted \ce{Ga2O3} layer.

To identify microscopic configurations compatible with the observed
spectral line shape, the experimental spectra were compared with
first-principles-based XANES simulations for several candidate nitrogen
configurations, see Figure~\ref{fig:1}~(a). 
The tested models included:
(i) substitutional nitrogen on oxygen sites (\ce{N_{O}}, three curves
combined in the lower panel), (ii) interstitial \ce{N-N} paired
configurations (\ce{i5-i9}), representing positions adopted from \cite{7},
magenta dotted line, (iii) vacancy-assisted molecular configurations
(\ce{i5-i9-V_{O}}, green dotted line). The substitutional
\ce{N_{O}} models fail to reproduce the dominant $\pi^{*}$ resonance and do not
capture the experimental near-edge line shape. In contrast, molecular
configurations reproduce both the intense $\pi^{*}$ resonance and the
characteristic pre-$\pi^{*}$ spectral shoulder. Within experimental
sensitivity, the data therefore constrain the dominant nitrogen state in
the implanted layer to \ce{N2}-like configurations, although a minor
contribution from non-molecular motifs cannot be excluded within the
intermediate thermal window.

The combination of temperature-dependent N~$K$-edge XANES and
first-principles modeling reveals that nitrogen introduced by ion
implantation in (100) \ce{\beta-Ga2O3} does not
predominantly incorporate as substitutional \ce{N_{O}} defects but
instead self-organizes into \ce{N-N} bonded molecular configurations
stabilized by implantation-induced disorder. While substitutional
nitrogen has long been considered the most likely configuration for
nitrogen doping in gallium oxide \cite{6,7}, direct spectroscopic
verification of this assumption has remained limited, but our
spectroscopic results indicate that molecular nitrogen plays the
dominant role in the implanted and annealed gallium oxide.

A key insight emerges from the structural relaxation of the candidate defect configurations. 
In the isolated \ce{N2} model calculation, the optimized \ce{N-N} bond length is \qty{1.094}{\angstrom}, in excellent agreement with the experimental bond length of the free nitrogen molecule (\qty{\approx 1.10}{\angstrom}). 
This confirms that the computational framework accurately reproduces the intrinsic molecular geometry of nitrogen. 
When the \ce{N-N} pair is embedded in the oxide sub-lattice (our case oxygen vacancy), however, the bond length increases to \qty{1.165}{\angstrom}.
Such elongation is characteristic of partial occupation of antibonding $\pi^{*}$ orbitals and indicates that the molecule is weakly charged through
interaction with the surrounding lattice. 
In molecular terms, the configuration can therefore be described as a partially reduced species \ce{N2^{\delta-}} stabilized by the defect environment.

The Bader charge analysis provides direct insight into the electronic
origin of this bond elongation. For substitutional nitrogen on the
oxygen site (\ce{N_{O}}), the calculated charge corresponds to
approximately -1\textit{e} on the nitrogen atom. This reflects the ionic
character of the \ce{Ga-N} bond and is consistent with the expected
electronic configuration of substitutional nitrogen in an oxide lattice.
In contrast, the interstitial \ce{N-N} configuration (\ce{i5-i9}) yields nearly
neutral nitrogen atoms, with the total charge of the \ce{N2} unit remaining
close to zero. This behavior is typical of a covalently bonded molecular
unit embedded in the lattice rather than two independent atomic defects.
A different situation emerges when an oxygen vacancy is present in the
vicinity of the nitrogen pair. In the \ce{i5-i9-V_{O}} configuration
both nitrogen atoms acquire additional electronic density, leading to
partial negative charging of the molecular unit. Oxygen vacancies in
oxides are well known to act as electron donors, and the calculated
Bader charges confirm that a fraction of this donated charge populates
the antibonding orbitals of the nitrogen molecule. The resulting charge
state weakens the \ce{N-N} bond and naturally explains the calculated bond
elongation relative to the neutral molecule. The simultaneous increase
in bond length and accumulation of electronic density on the molecular
unit therefore provides a consistent structural and electronic
fingerprint of vacancy-stabilized molecular nitrogen.

The molecular character of the \ce{N-N} complexes is further reflected in the calculated charge-transition levels shown in Figure~\textbf{S1}. 
For the interstitial pair configuration (\ce{i5-i9}), the calculated transitions occur at approximately $E_{V}+0.68$ and $E_V + \qty{1.63}{\eV}$, while the vacancy-assisted configuration (\ce{i5-i9-V_{O}}) exhibits transitions at $E_{V}+0.21$ and $E_V + \qty{3.78}{\eV}$. 
These levels span a broad portion of the band gap and demonstrate that the \ce{N-N} complex can stabilize several charge states. 
Such behavior is characteristic of molecular defects whose electronic structure is governed by the filling of bonding and antibonding orbitals associated with the \ce{N-N} unit. 
For comparison, hybrid-DFT calculations performed
for substitutional \ce{N_{O}} in \ce{\beta-Ga2O3}
reveal a distinctly different electronic behavior. The calculated
thermodynamic transition levels form a compact sequence close to the
valence-band edge. 
In particular, the $(+2/+1)$ and $(+1/0)$ transitions occur within approximately \qtyrange{0.6}{1.6}{\eV} above the VBM, while the $(0/-1)$ transition lies deeper in the gap at roughly \qtyrange{2}{3.6}{\eV} depending on the specific oxygen site. 
Such a hierarchy of charge states is characteristic of a localized atomic acceptor whose electronic structure is primarily governed by N~2p states hybridized with the valence band of the oxide lattice. 
In contrast, the \ce{N-N} configurations considered in the present work exhibit thermodynamically accessible charge states distributed over a much broader energy range across the band gap (we will return to this below). 
This behavior is naturally expected for molecular complexes, where the electronic structure is controlled by the occupation of molecular orbitals associated with the \ce{N#N} bond rather than by a single impurity level.

Importantly, the structural analysis demonstrates that the formation of molecular nitrogen does not require the presence of a stable \ce{\gamma-Ga2O3} phase. 
X-ray diffraction
patterns (Figure~\textbf{S2}) show that implantation initially induces a
metastable \ce{\gamma} component in the damaged region, while annealing restores
long-range \ce{\beta}-phase order. The persistence and strengthening of the
molecular spectral signature even after the disappearance of the \ce{\gamma}
diffraction features indicate that nitrogen molecularization is governed
by local defect environments rather than by long-range polymorphic
stability. In this sense, the implantation-induced defect landscape
provides the structural conditions necessary for stabilizing \ce{N-N} bonded
units within the oxide lattice. A comparison of the \ce{\beta} and \ce{\gamma} crystal
structures provides an additional structural perspective on this
stabilization mechanism. Several lattice positions characteristic of the
\ce{\gamma} polymorph correspond closely to interstitial sites within the
\ce{\beta-Ga2O3} framework \cite{29}. Ion
implantation therefore creates a defect landscape in which local atomic
arrangements temporarily reproduce \ce{\gamma}-like coordination motifs inside the
\ce{\beta} lattice. This observation indicates that the stabilization of \ce{N2} units
is governed primarily by local defect topology rather than by the
long-range crystallographic identity of the host phase. Even after the \ce{\gamma}
diffraction signatures disappear, implantation-induced vacancy clusters
and distorted coordination motifs can remain locally preserved and
continue to act as trapping environments for molecular nitrogen.

Within such environments, enlarged interstitial volumes and
under-coordinated motifs can trap nitrogen atoms. When two nitrogen
atoms occupy adjacent positions within these motifs, the formation of an
\ce{N-N} bond becomes energetically favorable (see Figure~\textbf{S1}),
leading to the stabilization of molecular \ce{N2}-like complexes within the
\ce{\beta-Ga2O3} matrix. Even after annealing
restores long-range \ce{\beta}-phase order, remnants of these local defect
topologies may persist and act as metastable trapping sites for
molecular nitrogen. Raman measurements independently confirm the
presence of molecular nitrogen in the samples (data not shown). RBS/c
results (Figure~\textbf{S3}) show that, after annealing, the
defect-related signal remains confined to the near-surface implanted
region, although its profile deviates from the as-implanted SRIM
distribution, indicating thermally driven defect redistribution. While
outward nitrogen migration cannot be excluded based on RBS/c alone,
which cannot detect nitrogen directly, the ``Relative Defect
Concentration'' reflects defect accumulation rather than nitrogen
content and therefore does not provide direct evidence of nitrogen
diffusion. In contrast, the SRIM-predicted nitrogen profile for RT
implantation is in very good agreement with the SIMS data
(Figure~\ref{fig:1}(c)). Crucially, the SIMS comparison before and after
annealing (Figure~\ref{fig:1}(d)) reveals a clear shift of nitrogen
toward the surface, providing direct evidence of diffusion toward the
near-surface region upon thermal treatment.

It should be noted that first-principles calculations for equilibrium
defect formation in \ce{\beta-Ga2O3} predict
relatively high formation energies for isolated interstitial \ce{N2} species.
However, ion implantation represents a strongly nonequilibrium process
that generates a high density of vacancies, interstitials, and defect
clusters. Such defect-rich environments can substantially modify the
local thermodynamic landscape and provide metastable trapping sites
where molecular nitrogen configurations become energetically stabilized.
Additional support for a molecular interpretation of nitrogen-related defects comes from electrical defect spectroscopy studies of nitrogen-implanted \ce{\beta-Ga2O3}. 
Deep-level
optical spectroscopy has revealed a prominent nitrogen-related trap
located approximately $E_{C}-\qty{2.9}{eV}$ below the conduction
band. Remarkably, this defect is characterized by an unusually large
Franck-Condon relaxation energy of about \qty{1.4}{\eV}, indicating extremely
strong lattice relaxation accompanying the electronic transition
\cite{1,30}. Such a large Franck-Condon shift is exceptional even among
WBG semiconductors and implies a defect configuration undergoing
substantial structural reorganization during charge capture or emission.
Molecular defect complexes provide a natural microscopic explanation for
this behavior. Changes in the charge state of an \ce{N-N} unit modify the
occupation of antibonding $\pi^{*}$ orbitals and directly alter the \ce{N-N} bond
length, thereby inducing significant relaxation of the surrounding
lattice. This strong coupling between electronic occupation and
molecular geometry offers a natural explanation for the unusually large
relaxation energy reported for nitrogen-related trap states.

Taken together, the structural, electronic, and spectroscopic data suggest a heterogeneous nitrogen speciation during the thermal evolution of the implanted layer. 
At lower temperatures, the system likely contains a heterogeneous mixture of configurations, including substitutional \ce{N_{O}} defects predicted by first-principles calculations as deep acceptors, interstitial \ce{N-N} pairs (\ce{i5-i9}), and vacancy-assisted molecular complexes (\ce{i5-i9-V_{O}}). 
The coexistence of these configurations is consistent with the strongly nonequilibrium nature of ion implantation and the broad distribution of defect environments created in the damaged layer. 
As the material is annealed, defect recombination and structural relaxation progressively favor the stabilization of molecular nitrogen species. 
Consequently, after annealing at temperatures approaching \qty{725}{\celsius}, the nitrogen population becomes dominated by \ce{N2}-like configurations trapped within the oxide matrix.

This mechanism provides a natural explanation for the well-known difficulty of achieving effective $p$-type doping in gallium oxide under ion implantation conditions. 
In this framework, nitrogen introduced via implantation does not behave as a simple electrically inactive impurity; instead, it self-organizes into molecular-like complexes that are electronically decoupled from the valence band, thereby preventing the formation of shallow acceptor states. 
In this way, the implantation-driven incorporation pathway effectively bypasses the conventional substitutional doping channel.

While first-principles calculations predict substitutional nitrogen at
the oxygen site as the thermodynamic ground state, the conditions
associated with ion implantation deviate strongly from equilibrium. The
implantation process generates a dense population of defects and
introduces nitrogen atoms in close spatial proximity within collision
cascades. Under these conditions, nitrogen atoms can readily interact
and form \ce{N-N} bonded configurations. Once formed, such paired
configurations represent locally stable states that can persist during
subsequent thermal treatment, hindering the system from relaxing toward
the thermodynamically preferred substitutional configuration.

The present results show that ion implantation in (100)~\ce{\beta-Ga2O3} drives dopant molecularization, where nitrogen atoms preferentially form \ce{N2}-like configurations rather than isolated substitutional \ce{N_{O}} defects. 
Temperature-dependent N~$K$-edge XANES combined with first-principles modeling demonstrates that implanted nitrogen self-organizes into \ce{N-N} bonded molecular units stabilized within the defect-rich implantation layer. 
The dominant $\pi^{*}$ resonance observed in the N~$K$-edge spectra provides a direct fingerprint of the \ce{N#N} bond and evolves systematically toward the molecular \ce{N2} reference upon annealing.
Complementary calculations further show that substitutional nitrogen cannot reproduce this spectroscopic signature, whereas \ce{N2}-like configurations formed through interstitial pairing and vacancy-assisted trapping naturally account for the observed spectral line shape. 
Taken together, the structural, electronic, and spectroscopic evidence reveal a thermally driven molecularization pathway for nitrogen in implanted (100)~\ce{\beta-Ga2O3}. 
This molecular self-organization effectively removes nitrogen from the substitutional doping channel and provides a microscopic explanation for the persistent difficulty of achieving $p$-type doping in this material.

\section*{Acknowledgements} \label{ackn}
\textit{
    These studies were financially supported by the project UMO-2020/39/B/ST5/03580 funded by the National Science Centre (NCN) in Poland. 
    This publication was partially developed under the provision of the Polish Ministry and Higher Education project ``Support for research and development with the use of research infrastructure of the National Synchrotron Radiation Centre SOLARIS'' under contract no 1/SOL/2021/2. 
    We acknowledge SOLARIS Centre for access to the PIRX beamline, where the measurements were performed. 
    RBS/c measurements were carried out at IBC at the Helmholtz-Zentrum Dresden-Rossendorf e. V., funded by the EU’s Horizon 2020 programme under grant agreement No 824096. 
    Research partially was funded by Warsaw University of Technology within the Excellence Initiative: Research University (IDUB) programme. 
    Crystal wafers for the present study were prepared within the Bundesministerium für Bildung und Forschung (BMBF) project under Grant No. 16ES1084K. 
    The authors thank Andreas Popp from Leibniz-Institut für Kristallzüchtung for critical reading of the paper.
}

\section*{Declaration of interests}
\textit{
    The authors report no conflict of interest.
}

\bibliographystyle{unsrt}

\bibliography{references.bib}

@article{1,
  author = {Ghadi, H. and Cornuelle, E. and Mcglone, J. F. and Senckowski, A. and Sharma, S. and Wong, M. H. and Singisetti, U. and Ringel, S. A.},
  title = {
  	Comprehensive characterization of nitrogen-related defect states in \ce{\beta}-{Ga$_2$O$_3$} using quantitative optical and thermal defect spectroscopy methods
  },
  journal = {APL Materials},
  volume = {12},
  number = {9},
  pages = {091111},
  year = {2024}
}

@article{2,
  author = {Luan, S. and Dong, L. and Ma, X. and Jia, R.},
  title = {
  	The further investigation of {N}-doped \ce{\beta}-{Ga$_2$O$_3$} thin films with native defects for {S}chottky-barrier diode
  },
  journal = {Journal of Alloys and Compounds},
  volume = {812},
  pages = {152026},
  year = {2020}
}

@article{3,
  author = {Nikolskaya, A. and Okulich, E. and Korolev, D. and Stepanov, A. and Nikolichev, D. and Mikhaylov, A. and Tetelbaum, D. and Almaev, A. and Bolzan, C. A. and Buaczik, Jr., A. and Giulian, R. and Grande, P. L. and Kumar, A. and Kumar, M. and Gogova, D.},
  title = {Ion implantation in \ce{\beta}-{Ga$_2$O$_3$}: physics and technology},
  journal = {Journal of Vacuum Science \& Technology A},
  volume = {39},
  number = {3},
  pages = {030802},
  year = {2021}
}

@article{4,
  author = {Nakanishi, M. and Wong, M. H. and Yamaguchi, T. and Honda, T. and Higashiwaki, M. and Onuma, T.},
  title = {Effect of thermal annealing on photoexcited carriers in nitrogen-ion-implanted \ce{\beta}-{Ga$_2$O$_3$} crystals detected by photocurrent measurement},
  journal = {AIP Advances},
  volume = {11},
  number = {3},
  pages = {035237},
  year = {2021}
}

@article{5,
  author = {Azarov, A. and Galeckas, A. and Bektas, U. and Hlawacek, G. and Kuznetsov, A.},
  title = {Optical absorption and emission in nitrogen-implanted {Ga$_2$O$_3$} controlled by dynamic defect annealing},
  journal = {Advanced Optical Materials},
  volume = {14},
  pages = {e03595},
  year = {2026}
}

@article{6,
  author = {Peelaers, H. and Lyons, J. L. and Varley, J. B. and Van de Walle, C. G.},
  title = {Deep acceptors and their diffusion in {Ga$_2$O$_3$}},
  journal = {APL Materials},
  volume = {7},
  number = {2},
  pages = {022519},
  year = {2019}
}

@article{7,
  author = {Shokri, A. and Melikhov, Y. and Syryanyy, Y. and Demchenko, I. N.},
  title = {Hybrid Density Functional Theory Study on the Formation Energies of Donor and Acceptor {N} Impurities in $\beta$-{Ga$_2$O$_3$}"},
  journal = {Physica Status Solidi (b)},
  volume = {262},
  pages = {2400448},
  year = {2025}
}

@incollection{8chapter5,
	IGNOREauthor = {Wang, Mengen and Mu, Sai and Varley, Joel B. and Hwang, Jinwoo and Van de Walle, Chris G.},
	author = {Wang, M. and Mu, S. and Varley, J. B. and Hwang, J. and Van de Walle, C. G.},
	editor = {Speck, J.S. and Farzana, E.},
	isbn = {978-0-7354-2500-2},
	title = {Chapter 5: {D}efects in $\beta$-{Ga$_2$O$_3$}: theory and microscopic Studies},
	booktitle = {Ultrawide Bandgap $\beta$-{Ga$_2$O$_3$} Semiconductor: Theory and Applications},
	publisher = {AIP Publishing LLC},
	doi = {10.1063/9780735425033_005},
	url = {https://doi.org/10.1063/9780735425033_005},
	eprint = {https://pubs.aip.org/book/chapter-pdf/12805698/9780735425033_005.pdf},
	year = {2023},
}

@incollection{8chapter6,
	IGNOREauthor = {Varley, Joel B. and Van de Walle, Chris G. and Farzana, Esmat},
	author = {Varley, J. B. and Van de Walle, C. G. and Farzana, E.},
	editor = {Speck, J.S. and Farzana, E.},
	isbn = {978-0-7354-2500-2},
	title = {Chapter 6: {D}opants in $\beta$-{Ga$_2$O$_3$}: from Theory to Experiments},
	booktitle = {Ultrawide Bandgap $\beta$-{Ga$_2$O$_3$} Semiconductor: Theory and Applications},
	publisher = {AIP Publishing LLC},
	doi = {10.1063/9780735425033_006},
	url = {https://doi.org/10.1063/9780735425033_006},
	eprint = {https://pubs.aip.org/book/chapter-pdf/12805716/9780735425033_006.pdf},
	year = {2023},
}

@article{9,
  author = {Wong, M. H. and others},
  title = {Acceptor doping of \ce{\beta}-{Ga$_2$O$_3$} by {Mg} and {N} ion implantations},
  journal = {Applied Physics Letters},
  volume = {113},
  number = {10},
  pages = {102103},
  year = {2018}
}

@inproceedings{10,
  author = {De Santi, C. and Fregolent, M. and Buffolo, M. and Higashiwaki, M. and Meneghesso, G. and Zanoni, E. and Meneghini, M.},
  title = {Deep levels and conduction processes in nitrogen-implanted {Ga$_2$O$_3$} {S}chottky barrier diodes},
  booktitle = {Oxide-based Materials and Devices XIII},
  series = {Proc. SPIE},
  volume = {12002},
  pages = {1200209},
  year = {2022}
}

@article{11,
  author = {Fons, P. and Tampo, H. and Kolobov, A. V. and Ohkubo, M. and Niki, S. and Tominaga, J. and Carboni, R. and Boscherini, F. and Friedrich, S.},
  title = {Direct Observation of Nitrogen Location in Molecular Beam Epitaxy Grown Nitrogen-Doped \ce{ZnO}},
  journal = {Physical Review Letters},
  volume = {96},
  number = {4},
  pages = {045504},
  year = {2006}
}

@article{12,
  author = {Schauries, D. and Ney, V. and Nayak, S. K. and Entel, P. and Guda, A. A. and Soldatov, A. V. and Wilhelm, F. and Rogalev, A. and Kummer, K. and Yakhou, F. and Ney, A.},
  title = {Incorporation of nitrogen in \ce{Co}:\ce{ZnO} studied by x-ray absorption spectroscopy and x-ray linear dichroism},
  journal = {Physical Review B},
  volume = {87},
  number = {12},
  pages = {125206},
  year = {2013}
}

@article{13,
  author = {Bazioti, C. and Azarov, A. and Johansen, K. M. and Svensson, B. G. and Vines, L. and Kuznetsov, A. Y. and Prytz, Ø.},
  title = {Role of Nitrogen in Defect Evolution in Zinc Oxide: {STEM}-{EELS} Nanoscale Investigations},
  journal = {The Journal of Physical Chemistry Letters},
  volume = {10},
  pages = {4725--4730},
  year = {2019}
}

@article{14,
  author = {Kato, Y. and Yamamoto, M. and Ozawa, A. and Kawaguchi, Y. and Miyoshi, A. and Oshima, T. and Maeda, K. and Yoshida, T.},
  title = {Analysis of Optical Properties and Structures of Nitrogen Doped Gallium Oxide},
  journal = {e-Journal of Surface Science and Nanotechnology},
  volume = {16},
  pages = {262--266},
  year = {2018}
}

@article{15,
  author = {Galazka, Z. and Uecker, R. and Klimm, D. and Irmscher, K. and Naumann, M. and Pietsch, M. and Kwasniewski, A. and Bertram, R. and Ganschow, S. and Bickermann, M.},
  title = {Scaling-up of bulk $\beta$-{Ga$_2$O$_3$} single crystals by the Czochralski method},
  journal = {ECS Journal of Solid State Science and Technology},
  volume = {6},
  pages = {Q3007},
  year = {2017}
}

@article{16,
  author = {Galazka, Z.},
  title = {Growth of bulk $\beta$-{Ga$_2$O$_3$} single crystals by the Czochralski method},
  journal = {Journal of Applied Physics},
  volume = {131},
  pages = {031103},
  year = {2022}
}

@article{17,
  author = {Turek, M. and Prucnal, S. and Droździel, A. and Pyszniak, K.},
  title = {Arc discharge ion source for europium and other refractory metals implantation},
  journal = {Review of Scientific Instruments},
  volume = {80},
  pages = {043304},
  year = {2009}
}

@article{18,
  author = {Turek, M. and Drozdziel, A. and Pyszniak, K. and Prucnal, S.},
  title = {Versatile plasma ion source with an internal evaporator},
  journal = {Nuclear Instruments and Methods in Physics Research Section B},
  volume = {269},
  pages = {700},
  year = {2011}
}

@article{19,
  author = {Kresse, G. and Hafner, J.},
  title = {
  	\textit{Ab initio} molecular-dynamics simulation of the liquid-metal-amorphous-semiconductor transition in germanium
  },
  journal = {Physical Review B},
  volume = {49},
  pages = {14251},
  year = {1994}
}

@article{20,
  author = {Kresse, G. and Furthmüller, J.},
  journal = {Computational Materials Science},
  title = {Efficiency of \textit{ab-initio} total energy calculations for metals and semiconductors using a plane-wave basis set},
  volume = {6},
  pages = {15},
  year = {1996}
}

@article{21,
  author = {Kresse, G. and Furthmüller, J.},
  title = {Efficient iterative schemes for \textit{ab initio} total-energy calculations using a plane-wave basis set},
  journal = {Physical Review B},
  volume = {54},
  pages = {11169},
  year = {1996}
}

@article{22,
  author = {Kresse, G. and Joubert, D.},
  title = {From ultrasoft pseudopotentials to the projector augmented-wave method},
  journal = {Physical Review B},
  volume = {59},
  pages = {1758},
  year = {1999}
}

@article{23,
  author = {Heyd, J. and Scuseria, G. E. and Ernzerhof, M.},
  title = {Hybrid functionals based on a screened {C}oulomb potential},
  journal = {The Journal of Chemical Physics},
  volume = {118},
  pages = {8207},
  year = {2003}
}

@article{24,
  author = {Xiao, W.-Z. and Wang, L.-L. and Xu, L. and Wan, Q. and Pan, A.-L.},
  title ={Electronic structure and magnetic properties in Nitrogen-doped $\beta$-\ce{Ga2O3} from density functional calculations},
  journal = {Solid State Communications},
  volume = {150},
  pages = {852},
  year = {2010}
}

@article{25,
  author = {Blanco, M. A. and Sahariah, M. B. and Jiang, H. and Costales, A. and Pandey, R.},
  title = {Energetics and migration of point defects in \ce{Ga2O3}},
  journal = {Physical Review B},
  volume = {72},
  pages = {184103},
  year = {2005}
}

@article{26,
  author = {Kyrtsos, A. and Matsubara, M. and Bellotti, E.},
  title = {On the feasibility of $p$-type \ce{Ga2O3}},
  journal = {Applied Physics Letters},
  volume = {112},
  pages = {032108},
  year = {2018}
}

@article{27,
  author = {Shokri, A. and Melikhov, Y. and Syryanyy, Y. and Demchenko, I. N.},
  title = {Point Defects in Silicon-Doped $\beta$-\ce{Ga2O3}: Hybrid-{DFT} Calculations},
  journal = {ACS Omega},
  volume = {8},
  number = {46},
  pages = {43732-43738},
  year = {2023},
  doi = {10.1021/acsomega.3c05557},
  URL = { https://doi.org/10.1021/acsomega.3c05557},
  eprint = { https://doi.org/10.1021/acsomega.3c05557	},
}

@article{28,
  author = {Joly, Y.},
  title = {X-ray absorption near-edge structure calculations beyond the muffin-tin approximation},
  journal = {Physical Review B},
  volume = {63},
  pages = {125120},
  year = {2001}
}

@unpublished{29,
  author = {Syryanyy, Y. and Domagala, J. Z. and Azarov, A. and Melikhov, Y. and Shokri, A. and Minikayev, R. and Liubchenko, O. and Zając, M. and Munnik, F. and Chernyshova, M. and Galazka, Z. and Krajczewski, J. and Ziętala, M. and Kuznetsov, A. and Demchenko, I. N.},
  title = {Silicon Implantation as a Route to Polymorph Engineering in {Ga$_2$O$_3$} (100)},
  note = {Submitted to J. of Phys. Chem. Lett.},
  year = {2026}
}

@article{30,
  IGNOREauthor = {Ghadi, Hemant and McGlone, Joe F. and Jackson, Christine M. and Farzana, Esmat and Feng, Zixuan and Bhuiyan, A. F. M. Anhar Uddin and Zhao, Hongping and Arehart, Aaron R. and Ringel, Steven A.},
  author = {Ghadi, H. and McGlone, J.F. and Jackson, C.M. and Farzana, E. and Feng, Z. and Bhuiyan, A.F.M. Anhar Uddin and Zhao, H. and Arehart, A.R. and Ringel, S.A.},
  title = {Full bandgap defect state characterization of \ce{\beta}-{Ga$_2$O$_3$} grown by metal organic chemical vapor deposition},
  journal = {APL Materials},
  volume = {8},
  pages = {021111},
  year = {2020},
	issn = {2166-532X},
	doi = {10.1063/1.5142313},
	url = {https://doi.org/10.1063/1.5142313},
	eprint = {https://pubs.aip.org/aip/apm/article-pdf/doi/10.1063/1.5142313/14564001/021111_1_online.pdf},
}

@article{34,
  author = {Hitchcock, A. P. and Brion, C. E.},
  title = {${K}$-shell excitation spectra of {CO}, \ce{N2} and \ce{O2}},
  journal = {Journal of Electron Spectroscopy and Related Phenomena},
  volume = {18},
  pages = {1--21},
  year = {1980}
}

\clearpage
\vspace*{5cm}
\begin{center}
	\Huge \textbf{Supplementary Material}
\end{center}
\clearpage

\includepdf[pages=-]{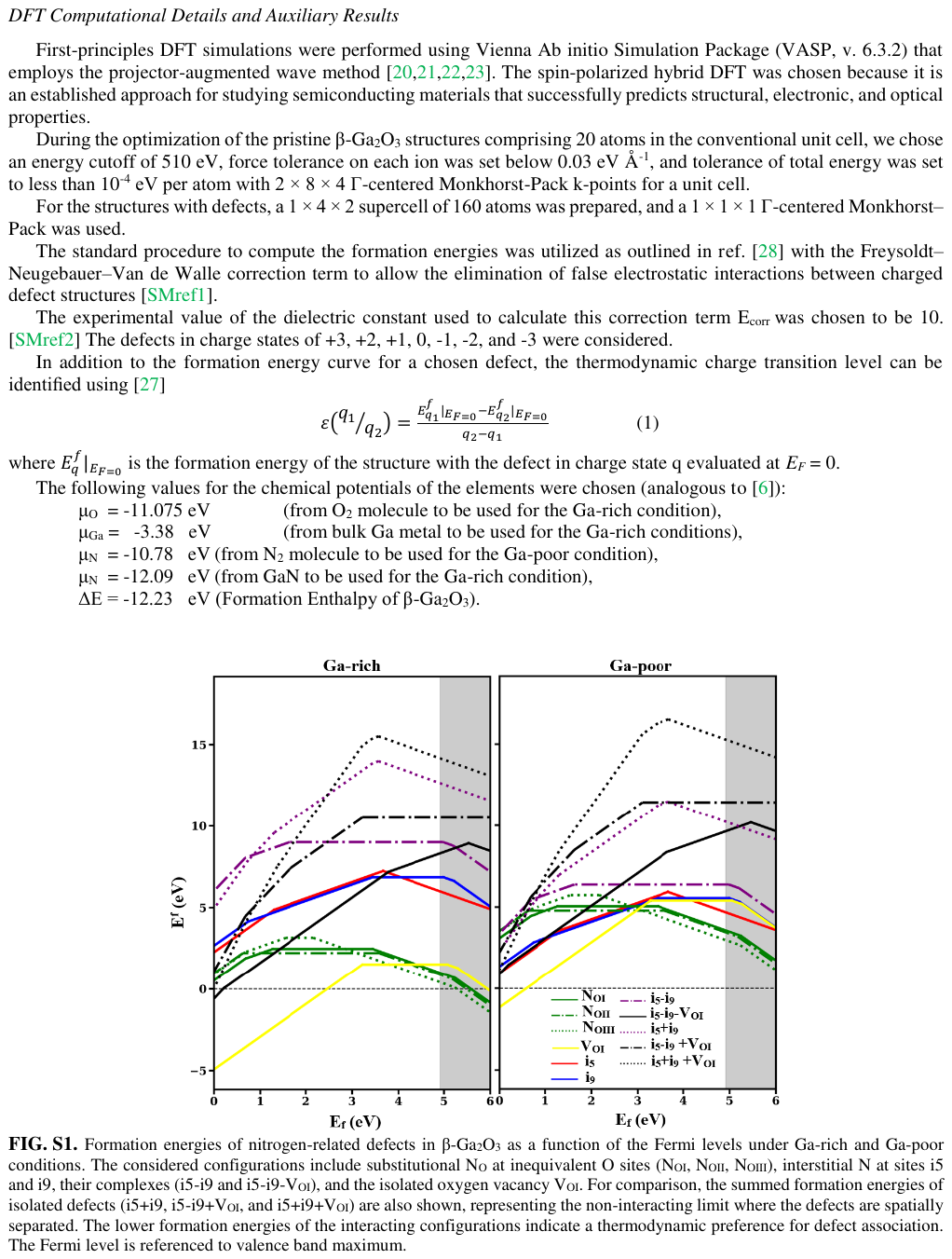}

\end{document}